\documentstyle[twoside,fleqn,espcrc2]{article}

\def\be{\begin{equation}}
\def\ee{\end{equation}}
\def\bea{\begin{eqnarray}}
\def\eea{\end{eqnarray}}

\def\bt{\begin{table}}
\def\et{\end{table}}
\def\ra{\rightarrow}
\def\ln{\left<}
\def\rn{\right>}

\def\lq{\Lambda_{QCD}}
\def\bl{\bar \Lambda}

\def\lbb{\Lambda_b}
\def\lq{\Lambda_{QCD}}
\def\bl{\bar \Lambda}

\def\ra{\rightarrow}

\def\om{\omega}
\def\qq{\ln \bar q q \rn}
\def\al{\alpha}

\def\np{Nucl. Phys.~}
\def\pl{Phys. Lett.~}
\def\prl{Phys. Rev. Lett.~}
\def\pr{Phys. Rev.~}

\def\bb{\bibitem}

\title{On the short distance nonperturbative corrections in \\
heavy quark expansion}
\author{S. Arunagiri
\address{Department of Nuclear Physics, University of Madras, Guindy Campus,\\
Chennai 600 025, Tamil Nadu, INDIA}}
\begin{document}

\begin{abstract}
We study the corrections due to renormalons to the heavy hadron decay width.
The renormalons contribution estimated in terms of finite gluon mass based on the 
assumption that the gluon mass represents the short distance nonperturbative
effects in the standard OPE (and hence in the heavy quark expansion). We found
that the corrections are about 10\% of the leading decay rate. We point out 
the implications for the assumption of quark-hadron duality in heavy quark expansion.
\end{abstract}
\maketitle

\section{Introduction}

The divergence of the perturbation theory at large order brings in
an ambiguity to physical quantities specified at short distances.
According to the present understanding, the ambiguity is given
by a class of renormalon diagrams which are chain of
$n$-loops in a gluon line. The phenomenon is deeply connected with
the operator product expansion (OPE). The perturbative part of the OPE
receives the renormalon corrections \cite{zak}.
Since in the OPE the first power-suppressed nonperturbative term is absent
and the renormalon corrections constitute short distance
nonperturbative effect, 
they are more significant than large order
corrections. 

The phenomenology of the power corrections is thus significant also
for the heavy quark expansion (HQE) which describes the
inclusive decays of heavy hadrons by an expansion in the
inverse powers of the heavy quark mass, $m_Q$.
As the inclusive decay rate of heavy hadrons scales like
the fifth power of the heavy quark mass,
the power corrections arise due to momenta smaller than the
heavy quark mass. However, these IR renormalons would,
being nonperturbative effect, have greater influence in
the HQE prediction of quantities of interest.
These short distance nonperturbative effects
can be sought for explaining the smaller lifetime of $\Lambda_b$.
We should note that these power corrections to heavy quark decay rate
represents the breakdown of the quark-hadron duality. Therefore, it may shed
light on the working of the assumption of quark-hadron duality in
the heavy quark expansion.

In this talk, we present the study on the renormalon corrections to
heavy hadron decay rate at the leading order, 
assuming that the nonperturbative short distance corrections given by 
the gluon mass that is much larger than the QCD scale, 
$\lambda^2 \gg \Lambda_{QCD}^2$. We carry out the analysis for both $B$ meson
and $\lbb$ heavy baryon. Our study shows that the short distance
nonperturbative corrections to the baryon and the meson
differ by a small amount which is significant for the smaller lifetime of the $\Lambda_b$.
In both the cases, these duality violating corrections are of the order 
of 10\%.
In the next section, the significance of the renormalons contribution 
is elucidated. The estimation of the $\lambda^2$
value for $B$ and $\lbb$, as renormalons corrections, using QCD sum rules
is presented in section 3. In view of the predicted $\lambda^2$ values,
the inclusive decay widths and the implications for quark-hadron duality are
discussed in section 4, followed by concluding remarks in section 5.

\section{Power Corrections}
For the correlator of hadronic currents $J$:
\be
\Pi(Q^2) = i \int d^4x e^{iqx} \ln 0|T\{J(x)J(0)\}|0\rn
\ee
where $Q^2 = -q^2$, the standard OPE is expressed as
\bea
\Pi(Q^2) &\approx& [\mbox{parton model}] 
(1+a_1 \al + a_2 \al^2 + ....)\nonumber\\
&& + O(1/Q^4)
\eea
where the power suppressed terms are quark and gluon operators.
The perturbative series in the above equation can be rewritten as
\be
D(\al) = 1 + a_0 \al \sum_{n = 1}^\infty a_n \al^n \label{ddd}
\ee
where the term in the sum is considered to be the
nonperturbative short distance quantity and is given by a set of renormalon graphs.
It is studied by Chetyrkin {et al} \cite{chet} assuming that the short
distance tachyonic gluon mass, $\lambda^2$,
imitates the nonperturbative physics of the QCD. This, for the gluon
propagator, means:
\be
D_{\mu \nu}(k^2) = {\delta_{\mu \nu} \over {k^2}} \ra
\delta_{\mu \mu}\left( {1 \over {k^2}}+{\lambda^2 \over {k^4}}\right)
\label{gp}
\ee
On one hand, the nonperturbative short distance corrections are argued to be the
$1/Q^2$ correction in the OPE. On the other hand, the may have deep insight
of the confining configuration of the QCD vacuum.

Let us assume that
the gluon mass $\lambda^2 \gg \lq^2$ which
is not necessarily to be tachyonic one. Such a situation finds similarity in
the case of QED as well as QCD. For an $e^- e^+$ pair, separated by a
distance $r$, contained in a cage of
dimension $L$, $L \gg r$, the potential is of the form
\be
V_{e^- e^+}(r) = -{\alpha_e \over r} + \hbox{const.} \alpha_e {r^2 \over {L^3}}
\label{ve}
\ee
The power correction to the leading Coulomb term is can be interpreted as
the interaction of dipole with its images. This can also be obtained in terms
of one photon exchange, with the 
virtuality $\sim L^{-1}$. 
In the case of QCD, the heavy quark potential is given by
\be
V(r) = -{4\alpha(r) \over {3r}}+kr \label{v}
\ee
where $k \approx$ 0.2 GeV$^2$,
representing the string tension. Now, the correction term in (\ref{v})
can be considered to be due to one gluon exchange. If the gluon
happens to be massive one, we get the gluon propagator modified,
as given in (\ref{gp}). It has been argued in
\cite{bal} that the linear term can be replaced by a term of order $r^2$.
It is equivalent to replace $k$ by a term describing the ultraviolet
region.
For the potential in (\ref{v}),
\be
k \ra k^\prime \ approx constant \times \al \lambda^2
\ee
In replacing the coefficient of the term of $O(r)$ by $\lambda^2$,
we make it consistent by the renormalisation factor.
Thus the coefficient $\sigma(\lambda^2)$ is given by \cite{ani}:
\be
\sigma(\lambda^2) = \sigma(k^2)\left(\alpha(\lambda^2) \over
\alpha(k^2)\right)^{18/11} \label{rge}
\ee
Introduction of $\lambda^2$ brings in a small correction to the Coulombic term.
By use of (\ref{rge}), we specify the effect at both the ultraviolet
region
and the region characterised by the QCD scale.
Then, we rewrite (\ref{ddd}) as
\be
D(\al) = 1 + a_0 \al\left(1+{{k^\prime}^2 \over {\tau^2}}\right)
\label{dc}
\ee
where $\tau$ is some scale relevant to the problem and ${k^\prime}^2$ 
should be read
from (\ref{rge}). We would apply this to estimate the power correction
in the heavy quark expansion. 

We should note that in the QCD sum rules approach, the scale involved
in is given by the Borel variable which is about 0.5 GeV. But
in the heavy quark expansion the relevant scale is
the heavy quark mass, greater than the
hadronic scale. Thus, there it turns out to be infrared renormalons
effects.
But, still it represents the short distance nonperturbative
property, by virtue of the gluon mass being as high as the hadronic
scale.

\section{Heavy-Light Hadrons}
\subsection{$B$ Meson} For the heavy light current, $J(x) = \bar Q(x) i\gamma_5
q(x)$, the QCD sum rules for $B$ meson is already known \cite{ben}:
\bea
{\tilde f}_B^2 e^{-\bl_B/\tau} &=& {3 \over {\pi^2}}
\int_0^{\om_c} d \om \om^2 e^{-\om/\tau}D(\al)_B \nonumber\\
& &- \qq + {1 \over {16 \tau^2}}\ln g \bar q \sigma G q \rn +...\label{hl}
\eea
where $\om_c$ is the duality interval, $\tau$ the Borel variable, $\bl_B$ 
the mass gap parameter, the values of condensates given elsewhere below and
\be
D(\al)_B = 1 + {a_B \al}\left[1+{\lambda^2 \over {\tau^2}}
\left({\alpha(\lambda^2) \over {\alpha(\tau^2)}}\right)^{-18/11}\right]
\ee
where $a_B =   17/3+4\pi^2/9-4$log$(\om/\mu)$, with $\mu$
is chosen to be 1.3 GeV.

With the duality interval of about 1.2-1.4 GeV which is little smaller than
the
onset of QCD which corresponds to 2 GeV and $\bl \geq$ 0.6 GeV,
we get 
\be
\lambda^2 = 0.35~\mbox{GeV$^2$}. \label{l2m}
\ee

\subsection{$\lbb$ Baryon}
For the heavy baryon current
\be
j(x) = \epsilon^{abc}(\bar q_1(x)C \gamma_5 t q_2(x))Q(x)
\ee
where $C$ is charge conjugate matrix, $t$ the antisymmetric flavour matrix
and $a, b, c$ the colour indices, the QCD sum rules is given \cite{dai} by
\bea
{1 \over 2} f_{\Lambda_b}^2 e^{\bl/\tau} &=&
{1 \over {20\pi^4}}\int_0^{\om_c} d \om \om^5
e^{-\om/\tau}D(\al)_{\lbb} \nonumber\\
&& + {6 \over {\pi^4}}E_G^4 \int_0^{\om_c} d \om e^{-\om/\tau} \nonumber\\
&&+{6 \over {\pi^4}}E_Q^6e^{-m_0^2/8\tau^2}
\eea
where 
\be
D(\al)_{\lbb} = 1-{\al \over {4\pi}}a_{\lbb} \left(1+{\lambda^2 \over {\tau^2}}\right)
\ee
with $a_{\lbb} = r_1$log$(2\om/\mu)-r2)$. With
$f_{\Lambda_b}^2$ = $0.2 \times 10^{-3}$~GeV$^6$,
$\ln \bar q q \rn  = -0.24^3$ GeV$^3$, $\ln g \bar q \sigma G q \rn =
m_0^2 \ln \bar q q \rn$, $m_0^2 = 0.8$ GeV$^2$, $\ln \al GG \rn
= 0.04$ GeV$^4$.
As in the meson case, we obtain 
\be
\lambda^2 = 0.4~\mbox{GeV$^2$}.
\ee

In both the cases above, the gluon mass turn out to be about 0.6 GeV and above.
They mean a somewhat large coefficient of the term at large order in the 
perturbative expansion.

\section{Inclusive Decays and Quark-Hadron Duality}
According to HQE, the inclusive decay
rate of a weakly decaying heavy hadron is, at the leading order, given by
\be
\Gamma(H_b) = \Gamma_0
\left[1-{\al \over \pi}\left( {2 \over 3}g(x) - \xi \right) \right] \label{rt}
\ee
where $\xi$ stands for the renormalons corrections:
\be
\Delta \Gamma(H_Q)_{IR} \approx a_0 \alpha_s \sqrt{{\lambda^2 
\over {m_Q^2}}}\left({\alpha(\lambda^2) \over {\alpha(m_Q^2)}} \right)^
{-9/11}
\ee
Numerically, the IR renormalon corrections are found to be
\bea
\Delta \Gamma (B)_{IR} & \approx & 0.1 \Gamma_0\\
\Delta \Gamma (\lbb)_{IR} & \approx & 0.11 \Gamma_0
\eea
where $\Gamma_0$ $b$-quark decay rate at the tree level:
\be
\Gamma_0 = {G_f^2 |V_{KM}|^2 m_b^5 \over {192 \pi^3}} f(x)
\ee
The corrections being about 10\% signify that the decay width is perturbatively under 
control. On the other hand, these corrections arise due to nonperturbative
physics at short distance.

The assumption on the gluon mass has hence heuristic meaning. Though it 
is used to evaluate the renormalons effects, this would mean physics of
confining configurations quantitatively. As is well known, quark-hadron duality
signify the interplay of confinement and asymptotic freedom at a particular
kinematic regime. Thus, the above quantitative measure can be construed to be
duality violating effects.

In HQE, it has been pointed out in \cite{mas} 
that the violation of duality in HQE is of exponential/oscillating in 
nature:
\be
\Pi(Q^2)_{violation} = e^{-C Q^2/\lq^2}
\ee
where $C$ is constant and $Q^2$ is the energy scale. However, this violating
effect is not quantified. This violating quantity has been attributed to
the discrepancy in the inclusive properties as predicted by HQE.

On the other hand, in \cite{bmn}, the weak decay of heavy hadrons is studied
in the 't Hooft model. It has been found that the duality holds good with 
the presence of terms of order $1/m_Q$. Such a term is absent in the HQE. 
We should note that the first-power-suppressed term is absent in the OPE
itself. We should note that the 't Hooft QCD is 1+1 dimensional where
confinement is bulit-in. But, in QCD, the phenomenon of confinement is
not understood. Therefore, we cannot expect every aspect of the two-dimensional 
QCD to agree in {\it toto} with the QCD.

Recently, it has been shown in \cite{arun} that the four-quark operators are
indeed responsible for the discrepancy of lifetimes of $B$ and $\lbb$. If 
one assumes that the HQE is saturated by the terms up to three in $1/m_b$,
then the differences between the hadrons under $SU(3)_f$ symmetry yields
the expectation values of the four-quark operators of $B$-hadrons such the
the ratio, $\tau(\lbb)/\tau(B)$, close to the experimental value. This would
straightly mean that duality indeed holds good in HQE as far as the bottom sector
is concerned. On the other hand, the present study shows that if the renormalons
corrections are considered to be duality violating effects, then the violation
of duality is significantly few per cent.

\section{Conclusion}
Our assumption that the gluon mass, $\lambda^2 \gg \Lambda_{QCD}^2$, that
imitates the nonperturbative physics at short distance. This signifies
some unknown confining effects that is given by duality breaking effects in
the standard OPE and hence in HQE. We found that these effects are about
10\% of the leading decay rate.
Our studies show that the inclusive decays of heavy hadrons can be studied within
the framework of the HQE, notwithstanding the aspects like exponential violation 
\cite{cds,bgl1} which have not been quantified.

Use of constraints of the mass gap parameter due to the kinetic energy term
and the present value on the difference in the mass gap parameter of $B$ and $\lbb$
would result in more precision of the duality braking effects. Besides, such studies
in the charm sector are also relevant.

\section{Acknowledgements}
The author is grateful to Prof. Y. Okada for discussions and hospitality
at the Theory Group, KEK, Japan where a part of this work culminated,
Prof. S. Narison for discussions and Prof. V. I. Zakharov for useful 
communications. He thanks Prof. Apoorva Patel and the symposium organisers
for invitation to the wonderful event and warm hospitality during the
symposium.
{\em The author attains pleasure in dedicating this work to Prof. T. Nagarajan,
former Head of the Department of Nuclear Physics, University of Madras, India.}

\end{document}